%% file: acl_main.tex
\documentclass[11pt]{article}

\usepackage[preprint]{acl}
\usepackage{times}
\usepackage{latexsym}
\usepackage[T1]{fontenc}
\usepackage[utf8]{inputenc}
\usepackage{microtype}

\usepackage{inconsolata}

\usepackage{graphicx}
\usepackage{multirow}
\usepackage{amsmath}
\usepackage{enumitem}
\usepackage{amssymb}

\usepackage{booktabs} 
\usepackage{caption}  
\usepackage{colortbl}
\usepackage[table,xcdraw]{xcolor} 
\usepackage{pifont}
\usepackage{tcolorbox}
\tcbuselibrary{skins, breakable} 
\usepackage{soul} 
\definecolor{promptframe}{RGB}{60, 60, 60}   
\definecolor{promptbg}{RGB}{248, 248, 248}  
\definecolor{prompttitle}{RGB}{230, 230, 230} 

\newtcolorbox{promptbox}[2][]{
    enhanced,
    colback=promptbg,
    colframe=promptframe,
    coltitle=white,          
    colbacktitle=promptframe, 
    title=\textbf{#2},        
    fonttitle=\bfseries\small,
    fontupper=\small\ttfamily, 
    boxrule=1pt,
    arc=2pt,
    left=5pt, right=5pt, top=5pt, bottom=5pt,
    #1
}
\definecolor{promptbgcolor}{RGB}{245,245,245} 
\definecolor{promptframecolor}{RGB}{100,100,100} 
\definecolor{hl_green_bg}{RGB}{225, 255, 225}
\definecolor{hl_red_bg}{RGB}{255, 235, 235}
\definecolor{rowgray}{gray}{0.92}   
\definecolor{rowblue}{RGB}{230, 240, 255} 
\definecolor{hl_green_bg}{RGB}{225, 255, 225}
\definecolor{hl_red_bg}{RGB}{255, 235, 235}
\definecolor{dark_green}{RGB}{0, 100, 0}
\definecolor{dark_red}{RGB}{180, 0, 0}
\newcommand{\hlg}[1]{%
  \sethlcolor{hl_green_bg}% 设置背景条颜色
  \textcolor{dark_green}{\textbf{\hl{#1}}}% 颜色和加粗包在外面
}

\newcommand{\hlr}[1]{%
  \sethlcolor{hl_red_bg}% 设置背景条颜色
  \textcolor{dark_red}{\textbf{\hl{#1}}}% 颜色和加粗包在外面
}
\title{VERDICT: Verifiable Evolving Reasoning with Directive-Informed Collegial Teams for Legal Judgment Prediction}

\author{
  \textbf{Hui Liao\textsuperscript{1}},
  \textbf{Chuan Qin\textsuperscript{2}},
  \textbf{Yongwen Ren\textsuperscript{1}},
  \textbf{Hao Li\textsuperscript{3}},
  \textbf{Zhenya Huang\textsuperscript{1}},
  \textbf{Yanyong Zhang\textsuperscript{1}},
  \textbf{Chao Wang\textsuperscript{1}}
  \\
  \textsuperscript{1}University of Science and Technology of China \quad
  \textsuperscript{2}Chinese Academy of Sciences \quad
  \textsuperscript{3}iFLYTEK AI Research
  \\
  \texttt{\{liaohui2002, yovren\}@mail.ustc.edu.cn}, \texttt{chuanqin0426@gmail.com} \\
  \texttt{haoli5@iflytek.com}, \texttt{\{huangzhy, yanyongz, wangchaoai\}@ustc.edu.cn}
}

\begin{document}

\maketitle
\input{sections/0.abstract}

\input{sections/1.introduction1}

\input{sections/2.relatedwork}
\input{sections/4.method_new}
\input{sections/5.experiment}
\input{sections/6.conclusion}
\clearpage
\input{sections/limitations}
\input{sections/Ethics_Statement}

\raggedbottom
\bibliography{custom}

\input{sections/7.appendix}

\end{document}

%% file: sections/0.abstract.tex
\begin{abstract}
Legal Judgment Prediction (LJP) predicts applicable law articles, charges, and penalty terms from case facts. 
Beyond accuracy, LJP calls for intrinsically interpretable and legally grounded reasoning that can reconcile statutory rules with precedent-informed standards. However, existing methods often behave as static, one-shot predictors, providing limited procedural support for verifiable reasoning and little capability to adapt as jurisprudential practice evolves. We propose VERDICT, a self-refining collaborative multi-agent framework that simulates a \emph{virtual collegial panel}. VERDICT assigns specialized agents to complementary roles (e.g., fact structuring, legal retrieval, opinion drafting, and supervisory verification) and coordinates them in a traceable draft--verify--revise workflow with explicit Pass/Reject feedback, producing verifiable reasoning traces and revision rationales. To capture evolving case experience, we further introduce a Hybrid Jurisprudential Memory (HJM) grounded in the \textit{Micro-Directive Paradigm}, which stores precedent standards and continually distills validated multi-agent verification trajectories into updated Micro-Directives for continual learning across cases.
We evaluate VERDICT on CAIL2018 and a newly constructed CJO2025 dataset with a strict future time-split for temporal generalization. VERDICT achieves state-of-the-art performance on CAIL2018 and demonstrates strong generalization on CJO2025. To facilitate reproducibility and further research, we release our code and the dataset at \url{https://anonymous.4open.science/r/ARR-4437}.
\end{abstract}

%% file: sections/1.introduction1.tex
\section{Introduction}
% 法律判决预测（Legal Judgment Prediction, LJP），旨在基于案件事实自动预测法条、罪名和刑期，是智慧司法系统中的关键应用。它在现代法律版图中扮演着双重角色：一方面，它减轻了法院每年应对数千万立案的行政负担~\citep{spc2024report}，从而提升司法效率和一致性~\citep{cui2023survey}；另一方面，它促进了司法普惠（democratizes access to justice），使公民无需承担昂贵的法律咨询费用即可掌握潜在的案件结果~\citep{Feng2022epm}。然而，实现这些承诺要求系统具备严谨且可解释的裁决能力。由于法律逻辑固有的复杂性以及裁判尺度的动态调整，实现这一目标仍然是一个艰巨的挑战。
% Legal Judgment Prediction (LJP), which aims to automatically predict law articles, charges, and penalty terms based on case facts, serves as a pivotal application in intelligent judicial systems. It plays a dual role in the modern legal landscape: on one hand, it alleviates the administrative burden on courts grappling with tens of millions of filings annually ~\citep{spc2024report}, thereby enhancing judicial efficiency and consistency \citep{cui2023survey}; on the other hand, it democratizes access to justice, enabling citizens to grasp potential case outcomes without incurring expensive legal consultation costs ~\citep{Feng2022epm}. However, fulfilling these promises requires the system to possess rigorous and explainable adjudication capabilities. Due to the inherent complexity of legal logic and the dynamic adjustment of adjudication criteria, achieving this remains a formidable challenge.
Legal Judgment Prediction (LJP) predicts applicable law articles, charges, and penalty terms from case facts, and is increasingly used to support high-volume judicial workflows and public legal services~\citep{spc2024report,cui2023survey,Feng2022epm}. However, in this high-stakes setting, accuracy alone is insufficient: models must produce legally grounded and explainable predictions that align facts with constitutive elements and sentencing factors. This is difficult because judicial decisions must reconcile rigid statutory rules with evolving, context-dependent jurisprudential standards. Existing systems often rely on lexical shortcuts or statute-only retrieval, yielding decisions that are hard to justify when rules and standards diverge.

% 现有方法通常归为两类范式，但两者都面临显著的瓶颈。判别式模型，从早期的文本分类器~\citep{kim2014convolutional}到近期的图神经网络~\citep{yue2021neurjudge}，往往依赖高频文本模式而非深层法律推理。因此，它们在历史模式不适用的新发场景中泛化能力较差。另一方面，生成式大语言模型（LLMs）虽然具备强大的语义理解能力，却深受幻觉和缺乏严谨法理支撑的困扰。尽管检索增强生成（RAG）可以提供法理依据，但它缺乏像人类法官那样平衡二者的认知能力，无法弥合僵化的\textbf{“规则”}（法条）与灵活的\textbf{“标准”}（判例）之间的法理鸿沟~\citep{kaplow2013rules}。
% Existing approaches typically fall into two paradigms, yet both face significant bottlenecks. Discriminative models, ranging from early text classifiers ~\citep{kim2014convolutional} to recent graph neural networks \citep{yue2021neurjudge}, often rely on high-frequency textual patterns rather than deep legal reasoning. Consequently, they suffer from poor generalization in emerging scenarios where historical patterns do not apply. On the other hand, Generative Large Language Models (LLMs), while possessing strong semantic understanding, are plagued by hallucination and lack of rigorous legal grounding. Although Retrieval-Augmented Generation (RAG) can provide statutory grounds, it fails to bridge the jurisprudential gap between rigid \textbf{"Rules"} (statutes) and flexible \textbf{"Standards"} (precedents) \citep{kaplow2013rules}, lacking the cognitive capability to balance the two akin to human judges.
Existing approaches largely follow two paradigms, each with clear bottlenecks. Discriminative methods—ranging from dependency-aware models like TopJudge~\citep{zhong2018topjudge} to graph-interaction networks like LADAN~\citep{xu2020ladan}—often learn decision boundaries from frequent lexical and structural correlations. While effective on static benchmarks, they offer limited support for explicit fact-to-element alignment, making predictions brittle for novel fact patterns where such correlations are unreliable. Generative LLM-based approaches, such as LegalReasoner~\citep{shi-etal-2025-legalreasoner} or PLJP~\citep{wu2023pljp}, provide fluent case understanding and flexible reasoning, yet they may hallucinate or produce conclusions without verifiable legal support~\citep{huang2023lawyer}. Retrieval-Augmented Generation (RAG) partially mitigates this by grounding outputs in retrieved statutes~\citep{wu2023pljp}. However, it typically treats legal knowledge as a static repository, struggling to operationalize the evolving jurisprudential tension between rigid Rules (statutes) and flexible Standards (precedents)~\citep{kaplow2013rules}, especially when statutory text and case-based standards point to competing outcomes.

% 尽管近期已有探索，但当前的 LJP 研究仍受制于三个尚未解决的关键问题。首先，缺乏内在的可解释性。在法律领域，推理过程的可解释性与结果的准确性同样重要~\citep{bibal2021legal}，然而大多数模型作为“黑盒”运行，或忽视了构建对司法透明度至关重要的严谨、循序渐进的三段论推理链。此外，现有方法难以区分细粒度的法律细微差别。它们往往无法区分语义相似但在法律上截然不同的场景（例如“盗窃”与“侵占”），因为通用文本编码器依赖关键词匹配，而不是将复杂事实与具体的犯罪构成要件对齐。最后，当前系统缺乏动态经验积累。与通过实践成长的人类法官不同，现有方法将知识视为静态存储库，缺乏从判例中提炼中间智慧或适应法律标准持续演变的认知机制。
% Despite recent explorations, current LJP research remains constrained by three critical unresolved issues. Primarily, there is a lack of intrinsic interpretability. In the legal domain, the interpretability of the reasoning process is just as critical as the accuracy of the result \citep{bibal2021legal}, yet most models operate as "black boxes" or neglect the construction of rigorous, step-by-step syllogistic reasoning chains essential for judicial transparency. Furthermore, existing methods struggle to distinguish fine-grained legal nuances. They often fail to differentiate semantically similar but legally distinct scenarios (e.g., "Theft" vs. "Embezzlement") because generic text encoders rely on keyword matching rather than aligning complex facts with specific constitutive elements of crime. Ultimately, current systems suffer from the absence of dynamic experience accumulation. Unlike human judges who evolve through practice, existing methods treat knowledge as static repositories, lacking the cognitive mechanism to distill intermediate wisdom from precedents or adapt to the continuous evolution of legal standards.
Despite recent explorations, current LJP research remains constrained by three critical unresolved issues. First, intrinsic interpretability is still insufficient. In the legal domain, the transparency of the reasoning process is as crucial as outcome accuracy~\citep{bibal2021legal}. Yet, most models operate as "black boxes" or one-shot predictors, lacking the traceable, multi-stage deliberation process required to construct explainable reasoning chains. Second, models struggle to distinguish fine-grained legal nuances, particularly when semantic similarity diverges from legal logic. Generic encoders often conflate scenarios (e.g., "Theft" vs. "Embezzlement") based on surface-level textual overlap, failing to align facts with specific constitutive elements and boundary conditions essential for accurate qualification. Third, current systems suffer from the absence of dynamic experience accumulation. Unlike human judges who refine criteria through practice, most approaches treat knowledge as static, lacking the cognitive mechanism to distill abstract "Standards" into precise, evolving Micro-Directives that can be continuously updated from newly adjudicated cases.

To address these limitations, we present VERDICT, which integrates Directive-Informed memory with a Collegial Team of agents. VERDICT organizes judgment prediction as a traceable and explainable deliberation-and-verification workflow: the Court Clerk Agent extracts legally salient fact points; the Judicial Assistant Agent retrieves and filters applicable statutes and precedents; the Case-handling Judge Agent drafts a grounded opinion linking facts to legal elements; the Adjudication Supervisor Agent verifies the draft against statutes and a case-updated jurisprudential memory of precedent standards and Micro-Directives, and issues explicit Pass/Reject signals with corrective feedback to trigger revision; and the Presiding Judge Agent consolidates the verified draft into the final verdict. To more faithfully emulate how collegial panels reconcile statutory rules with precedent-informed discretion, we draw on the \textit{Micro-Directive Paradigm} from computational law~\citep{casey2016death}, which bridges rigid ``Rules'' and flexible ``Standards'' by distilling context-sensitive yet testable Micro-Directives. Building on this view, VERDICT incorporates a Hybrid Jurisprudential Memory (HJM) that stores precedent standards and evolving Micro-Directives, and continuously distills validated multi-agent verification trajectories into refined directives for continual learning across cases rather than static, one-shot inference.
\begin{itemize}[leftmargin=*, itemsep=2pt, parsep=0pt, topsep=2pt, partopsep=0pt]
\item We propose VERDICT (\textbf{V}erifiable \textbf{E}volving \textbf{R}easoning with \textbf{D}irective-\textbf{I}nformed \textbf{C}ollegial \textbf{T}eams), a self-refining multi-agent system that simulates a \emph{virtual collegial panel}  via a traceable draft--verify--revise loop, producing verifiable reasoning traces for judgment prediction.
\item We design a Hybrid Jurisprudential Memory (HJM) grounded in the \textit{Micro-Directive Paradigm}, which maintains precedent standards and evolving Micro-Directives and updates them by distilling verified multi-agent trajectories, enabling continual learning across cases.
\item We conduct comprehensive experiments on the widely used CAIL2018 benchmark and our newly constructed CJO2025 dataset. VERDICT achieves state-of-the-art performance on CAIL2018 and demonstrates strong temporal generalization on CJO2025.
\end{itemize}

%% file: sections/2.relatedwork.tex
\section{Related Work}
\subsection{Legal Judgment Prediction Paradigms}
% 法律判决预测（LJP）方法已从判别式分类演进为生成式推理。早期的判别式模型采用 CNN \citep{kim2014convolutional} 和 BERT \citep{devlin2019bert} 进行文本分类。为了捕捉逻辑层级，TopJudge \citep{zhong2018topjudge} 和 MPBFN \citep{yang2019mpbfn} 等依赖感知模型利用了有向无环图（DAGs）。后续工作，如 LADAN \citep{xu2020ladan}、NeurJudge \citep{yue2021neurjudge} 和 CTM \citep{liu2022ctm}，引入了图蒸馏和对比学习以区分细微的法律差别。然而，这些模型严重依赖高频模式 \citep{xiaocail2018}，导致在未见场景中泛化能力较差。
LJP approaches have evolved from discriminative classification to generative reasoning. Early discriminative models employed CNNs~\citep{kim2014convolutional} and BERT~\citep{devlin2019bert} for text classification. To capture logical hierarchies, dependency-aware models like TopJudge ~\citep{zhong2018topjudge} and MPBFN~\citep{yang2019mpbfn} utilized Directed Acyclic Graphs. Subsequent works, such as LADAN~\citep{xu2020ladan}, NeurJudge~\citep{yue2021neurjudge}, and CTM~\citep{liu2022ctm}, introduced graph distillation and contrastive learning to distinguish subtle legal nuances. However, these models rely heavily on high-frequency patterns~\citep{xiaocail2018}, leading to poor generalization in emerging scenarios.
% 近年来，大语言模型（LLMs）改变了这一范式。尽管 GPT-4 在考试中展现出潜力 \citep{katz2024gpt}，但在 LegalBench \citep{guha2023legalbench} 等基准测试上，通用大模型仍深受幻觉困扰 \citep{huang2023lawyer}。为增强推理能力，思维链（Chain-of-Thought） \citep{wei2022chain} 和 自修正（Self-Refine） \citep{madaan2023self, shinn2023reflexion} 等通用策略已被提出。具体到法律领域，LegalReasoner \citep{shi-etal-2025-legalreasoner} 引入逐步验证以纠正逻辑错误，而 ATRIE \citep{luo-etal-2025-automating} 利用检索来自动化法律概念解释。类似地，PLJP \citep{wu2023pljp} 将领域模型与检索增强生成（RAG）相结合。然而，这些方法主要依赖静态的内部知识或固定的检索语料库。它们缺乏人类法官那种\textit{动态的、基于经验的演化}机制，往往无法适应随时间变化裁判尺度
Recently, Large Language Models (LLMs) have shifted the paradigm. While GPT-4 shows promise in exams \citep{katz2024gpt}, generic LLMs on benchmarks like LegalBench~\citep{guha2023legalbench} still suffer from hallucinations \citep{huang2023lawyer}. To enhance reasoning, general strategies like Chain-of-Thought \citep{wei2022chain} and Self-Refine \citep{madaan2023self, shinn2023reflexion} have been proposed. Specific to law, LegalReasoner \citep{shi-etal-2025-legalreasoner} introduces step-wise verification to correct logical errors, while ATRIE \citep{luo-etal-2025-automating} utilizes retrieval to automate legal concept interpretation. Similarly, PLJP \citep{wu2023pljp} combines domain models with retrieval-augmented generation (RAG). However, these approaches primarily rely on static internal knowledge or fixed retrieval corpora. They lack the dynamic, experience-based evolution mechanism of human judges, often failing to adapt to the changing adjudication criteria over time.

\subsection{Multi-Agent Systems and Knowledge Evolution}
Multi-Agent Systems (MAS) solve complex tasks through role specialization. Frameworks like MetaGPT~\citep{hong2023metagpt} and ChatDev \citep{qian2024chatdev} utilize standardized operating procedures (SOPs), while AutoGen \citep{wu2024autogen} and AgentNet \citep{yang2025agentnet} enable decentralized coordination. A critical gap in legal MAS, however, is dynamic knowledge management. Existing memory modules like G-Memory \citep{zhang2025gmemory} treat interactions as static records rather than evolving wisdom. They fail to distill abstract ``Standard'' into precise directives over time. Our VERDICT addresses this by integrating a cognitive dual-layer memory into a ``Virtual Collegial Panel'', enabling agents to evolve their jurisprudential understanding through continuous practice.

%% file: sections/4.method_new.tex
\section{Method}
\begin{figure*}[t] 
    \centering
    \includegraphics[width=1.0\textwidth]{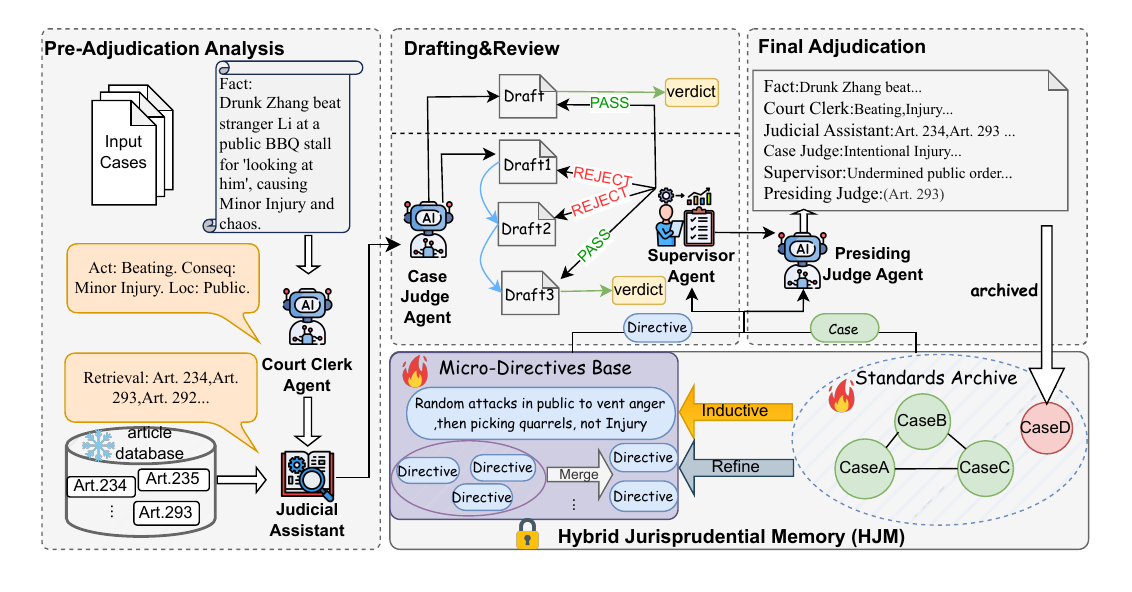} 
    \caption{The overall inference framework of VERDICT. It illustrates the interaction between the \textbf{Traceable Multi-Agent Workflow} and the \textbf{Hybrid Jurisprudential Memory (HJM)}. The process evolves through Preparation, Drafting \& Review, and Final Adjudication phases. Note that the Case Judge agent is instantiated using our domain-specific aligned expert model.}
    \label{fig:framework} 
\end{figure*}

In this section, we describe our proposed framework, VERDICT (Verifiable Evolving Reasoning with Directive-Informed Collegial Teams), as illustrated in Figure ~\ref{fig:framework}.
\subsection{Problem Formulation}
In this work, we focus on the task of Legal Judgment Prediction (LJP). Given a Legal Fact description formally defined as a token sequence $s_d = \{w_1^d, \ldots, w_{l_d}^d\}$ encompassing the case narrative, our objective is to predict the judgment result $j=(a, c, t)$. This target consists of three heterogeneous components: \textbf{Law Articles} $a \in Y_a$, representing specific articles from the Criminal Law; \textbf{Charges} $c \in Y_c$, defined by the constitutive elements of the crime; and \textbf{Imprisonment Terms} $t \in Y_t$, which are categorized into eleven distinct classes following standard conventions. Ultimately, the goal is to learn a mapping function $\mathcal{F}: s_d \to (a, c, t)$ to generate the accurate judicial reasoning and verdict based on the input facts.
% =================================================================
% 4.1 Workflow: 系统骨架
% =================================================================
\subsection{Traceable Multi-Agent Judicial Workflow}
\label{sec:workflow}
In real-world legal scenarios, Legal Judgment Prediction (LJP) is never an isolated classification task but a complex collaborative process covering case filing, research, drafting, deliberation, and final adjudication. To replicate this procedure, we design a multi-agent system simulating a real-world ``Virtual Collegial Panel.'' 

We formally define the judicial judgment prediction system as a collaborative framework based on a Directed Acyclic Graph (DAG), denoted as $\mathcal{M}=\langle\mathcal{U},\mathcal{A},\mathcal{S},\mathcal{P}\rangle$. Here, $\mathcal{U}$ represents the input space (i.e., the set of input cases); $\mathcal{A}$ is the set of agents; $\mathcal{S}$ denotes the intermediate states of the reasoning chain (e.g., drafts, feedback); and $\mathcal{P}$  governs the execution flow among nodes. Heterogeneous agents interact via a unified protocol, utilizing a context assembly function $\Psi(\cdot)$ to construct the prompt for each agent. We illustrate the detailed prompt designs for all agents in Appendix \ref{sec:appendix_prompts}. The inference process evolves through the following specialized roles:

\subsubsection*{Pre-Adjudication Analysis: Court Clerk Agent \& Judicial Assistant Agent}

Acting as the cornerstone of the workflow, the \textbf{Court Clerk Agent ($a_{clerk}$)} is responsible for extracting key factual points $o_{ext}$ (e.g., subjective criminal intent, specific criminal acts, and consequences) from the raw case dossier $u \in \mathcal{U}$. Subsequently, the \textbf{Judicial Assistant Agent ($a_{assist}$)} serves as the bridge to the Statutory Library $\mathcal{D}_{law}$, executing a two-stage retrieval process. It first obtains a Top-$K$ coarse candidate set $S_{vec}$ via dense vector retrieval based on $o_{ext}$, and then leverages the agent's robust semantic understanding to filter noise from the coarse set, identifying the reliable reference statute set $S_{statute}$:
\begin{gather}
    S_{vec} \leftarrow \mathrm{Search}(o_{ext}, \mathcal{D}_{law}, K), \\
    S_{statute} \sim \pi_{assist}(\cdot \mid \Psi(o_{ext}, S_{vec})).
\end{gather}

\subsubsection*{Drafting Phase: Case-handling Judge Agent}
This component functions as the core reasoning engine. Unlike generic LLMs, we employ the jurisprudentially aligned expert model $\pi^*_{\theta}$ (detailed in Sec.~\ref{sec:alignment}) to synthesize facts $o_{ext}$ and statutes $S_{statute}$. Crucially, this agent forms a refinement loop with the Supervisor. Let $h_t$ be the feedback from the previous round ($h_0$ is null), the judge generates a draft $y_{draft}^{(t)}$ utilizing its internalized legal logic:
\begin{equation}
    y_{draft}^{(t)} \sim \pi^*_{\theta}(\cdot \mid \Psi(o_{ext}, S_{statute}, h_{t})).
\end{equation}
Note that at this stage, the agent relies solely on model parameters and does not access the external memory $\mathcal{M}$.

\subsubsection*{Review Phase: Adjudication Supervisor Agent}

To ensure the judgment is not only legally valid but also appropriate, the \textbf{Adjudication Supervisor Agent ($a_{super}$)} intervenes. Unlike the drafter, this agent has access to the full Hybrid Jurisprudential Memory. Leveraging implicit knowledge from judicial precedents $S_{std}$ in the Contextual Standards Archive $\mathcal{M}_{std}$ and relevant Micro-Directives $S_{dir}$ in the Evolving Micro-Directive Base $\mathcal{M}_{dir}$, $a_{super}$ strictly scrutinizes the draft $y_{draft}^{(t)}$. If distinct discrepancies are found (e.g., compliant with statutes but violating a micro-directive on sentencing), it issues a rejection signal with corrective advice:
\begin{equation}
{\small
\begin{split}
\nonumber
\!\!\!(flag_t, fdbk_t) \! \sim\!  \pi_{super}(\cdot \!\mid\! \Psi(u, y_{draft}^{(t)}, 
\! S_{statute}, S_{std}, S_{dir})),
\end{split}
}
\end{equation}
here, $flag_{t} \in \{\mathrm{Pass}, \mathrm{Reject}\}$ serves as the judgment signal, and $fdbk_{t}$ provides natural language suggestions (e.g., "Incorrect charge qualification"). The system accumulates this feedback into the interaction history: $ h_t \leftarrow h_{t-1} \oplus fdbk_t.$
Subsequently, the workflow branches based on $flag_t$: a $\mathrm{Pass}$ signal (or reaching the maximum turn limit $T_{max}$) advances the verified draft to the \textit{Final Adjudication} phase, whereas a $\mathrm{Reject} $ triggers a redrafting iteration using the updated context $h_t$.
\subsubsection*{Final Adjudication Phase: Presiding Judge Agent}
Finally, the \textbf{Presiding Judge Agent ($a_{pres}$)} aggregates the refined draft and the comprehensive context to render the final explainable verdict with the complete reasoning process. As the ultimate decision-maker, $a_{pres}$ also possesses the capability to access the full memory bank to ensure global consistency, and performs case archiving:
\begin{equation}
{\small
\begin{split}
    \! \! \! y_{final} \! \sim\!  \pi_{pres}\big( \cdot \! \mid \!  \Psi( u, S_{statute}, S_{std},
    S_{dir}, y_{draft}^{(final)}) \big).
\end{split}
}
\end{equation}
\subsection{Domain-Specific Expert Alignment}
\label{sec:alignment}
To equip the Case-handling Judge with effective reasoning, we implement a two-stage alignment pipeline focusing on constructing logic-driven data.
\subsubsection{Protocol-Aware Instruction Tuning }
We first align the model with the protocol using a teacher model. By filtering inference results against ground truth, we retain accurate samples as the SFT set $\mathcal{D}_{train}$ to standardize output formats, while isolating erroneous predictions into an error set $\mathcal{D}_{fault}$. This splits the data into demonstrations for SFT and hard negatives for the next stage.
\subsubsection{Logic-Driven Contrastive Alignment}
To fix the logical hallucinations in $\mathcal{D}_{fault}$, we design an iterative correction mechanism. For an initial incorrect prediction (defined as the Loser $\hat{y}_l$) in $\mathcal{D}_{fault}$, a reflection model $M_R$ analyzes the logical gap and provides advice $r$. This guides the expert model $M_E$ to regenerate a legally valid response (defined as the Winner $\hat{y}_w$):
\begin{equation}
    \hat{y}_{w} = M_E(u, S, \hat{y}_l \oplus r).
\end{equation}
We collect successfully corrected trajectories to construct the preference dataset $\mathcal{D}_{pref}$, explicitly contrasting logic loopholes with reasoning:
\begin{equation}
    \!\!\!\mathcal{D}_{pref} \!=\! \left\{ \left( u, \hat{y}_w, \hat{y}_l \right) \!\mid\! \hat{y}_l \!\in \!\mathcal{D}_{fault}, \mathcal{V}(\hat{y}_w) \!=\! 1 \right\}.
\end{equation}
Where $\mathcal{V}(\cdot)$ is a validation function checking consistency with the statutes.
Finally, we apply standard Direct Preference Optimization (DPO)~\citep{rafailov2023dpo} on $\mathcal{D}_{pref}$ to sharpen the model's decision boundaries on these confusing legal concepts.
\subsection{The Hybrid Jurisprudential Memory (HJM) Mechanism}
\label{sec:memory}
To endow the system with continual learning and long-tail generalization, we construct the HJM, theoretically grounded in the ``Micro-Directive Paradigm''~\citep{casey2016death}. This architecture addresses the traditional jurisprudential dilemma between rigid ``Rule'' (clear but inflexible statutes) and vague ``Standard'' (flexible but noisy precedents). Casey and Niblett posit that AI can bridge this dichotomy by generating ``Micro-Directive''—precise norms possessing both the context-sensitivity of Standards and the ex-ante clarity of Rules. Inspired by this, our framework simulates the dynamic evolution from ``fuzzy Standard'' to ``precise Directives,'' distilling judicial experience into an intermediate modality between abstract law and concrete cases. Specifically, we operationalize this process by mapping empirical standards to the {Contextual Standards Archive} ($\mathcal{M}_{std}$) and instantiating evolved directives as the {Evolving Micro-Directive Base} ($\mathcal{M}_{dir}$). Formally, the memory is defined as $\mathcal{M} = \langle \mathcal{M}_{std}, \mathcal{M}_{dir}, \mathcal{D}_{law}, \Phi_{trans} \rangle$.
\subsubsection{Memory Structure and Retrieval}
The Contextual Standards Archive ($\mathcal{M}_{std}$): Constructed as an undirected graph $\mathcal{G}_{std}$, where each node $v_i=\langle\mathrm{txt}_i,\mathbf{h}_i,\Lambda_i,c_i\rangle$ represents an empirical ``Standard Precedent.'' Edges enforce intra-class consistency: $(v_i,v_j)\in\mathcal{E}\iff\mathbf{h}_i^\top\mathbf{h}_j \geq\tau\wedge(\Lambda_i\equiv\Lambda_j)$.
The Evolving Micro-Directive Base ($\mathcal{M}_{dir}$): Maintains dynamic units $m_{dir}=\langle r_{txt},\mathcal{S}_{conf},\mathcal{C}_{pos/neg},\Lambda_{anchor}\rangle$. Here, $r_{txt}$ offers precise, context-specific interpretation anchored to statute $\Lambda_{anchor}$, backed by confidence $\mathcal{S}_{conf}$ and supporting precedents.\\
\textbf{Multi-dimensional Retrieval.} 
To ensure jurisprudential relevance, we design a scoring mechanism for retrieving memory unit $m$ given the current case $u_{curr}$ and candidate statutes $S_{statute}$:
\begin{equation}
\label{eq:retrieval_score}
\begin{split}
\nonumber
    &\mathrm{Score}(m) \!=\! \alpha \cdot \mathrm{IoU}(\Lambda_m, S_{statute}) \\
    & \!+\! \beta \!\cdot\! \mathrm{Topo}(m, \mathcal{N}_{graph})
     \!+\! \gamma \!\cdot\! \mathrm{SemSim}(m, u_{curr}).
\end{split}
\end{equation}
% Here, IoU enforces consistency in statutory applicability; Topo indicates if $m$ lies on the activation diffusion path (signifying core knowledge); and SemSim captures factual proximity.
The score is a weighted sum of three \emph{metrics}: a statute/charge overlap metric $\mathrm{IoU}(\Lambda_m, S_{statute})$, a graph-topology metric $\mathrm{Topo}(m, \mathcal{N}_{graph})$, and a semantic-similarity metric $\mathrm{SemSim}(m, u_{curr})$. 
$\mathrm{IoU}$ measures the intersection-over-union between the tag set carried by $m$ and $S_{statute}$, enforcing legal applicability even under large factual variation.
$\mathrm{Topo}$ measures whether (and how strongly) $m$ is reached/co-activated along the activation-diffusion from $u_{curr}$, enabling associative recall of logically central but text-dissimilar precedents.
$\mathrm{SemSim}$ measures factual/lexical proximity between $m$ and $u_{curr}$ and serves as a surface-level anchor for retrieval. More details can be seen in Appendix~\ref{app:retrieval_score}.
\subsubsection{Evolutionary Mechanism: From Standards to Directives}
\begin{figure}[t]
    \centering
    \includegraphics[width=\columnwidth]{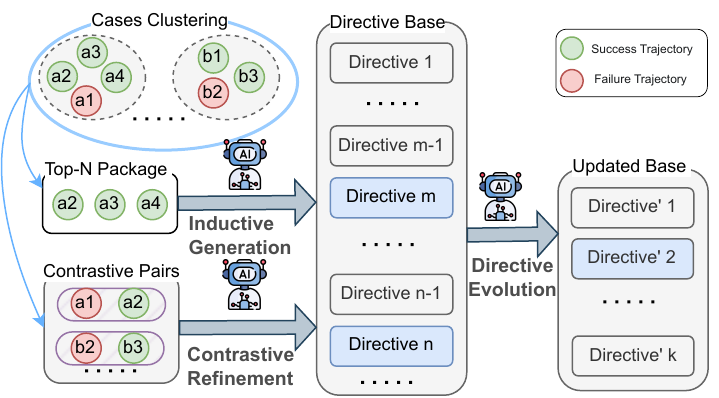}
    \caption{
    % The evolutionary mechanism of the Hybrid Jurisprudential Memory. It illustrates the lifecycle strategy $\Phi_{trans}$, transforming empirical Standards into precise Micro-Directives through ADD, REFINE, and PRUNE operations.
    The evolutionary mechanism of the Hybrid Jurisprudential Memory. The lifecycle strategy $\Phi_{trans}$ transforms archived empirical Standards into precise and compact Micro-Directives through three phases.}
    \label{fig:evolution}
\end{figure}
% As illustrated in Figure~\ref{fig:evolution}, we design a lifecycle strategy $\Phi_{trans}$ involving standard archiving and three-stage evolution \emph{(ADD, REFINE, PRUNE)}.
As illustrated in Figure~\ref{fig:evolution}, we design a lifecycle strategy $\Phi_{trans}$ consisting of standard archiving and a three-phase evolution process that (i) induces new Micro-Directives from consistent Standards, (ii) sharpens directive boundaries via contrastive evidence, and (iii) consolidates and removes directives to keep the memory compact.\\
\textbf{Standard Archiving.} 
Faithful recording of correct judicial practice ensures a pure experience pool. To capture the complete adjudication logic, we utilize a dense vector model $\mathcal{E}(\cdot)$ to encode the case facts $u$, the \textit{verified reasoning trace} $\tau$, and the verdict $y_{final}$ into persistent embeddings. 
\begin{equation}
\Phi_{archive}(u) = 
\begin{cases}
    \mathcal{E}(u \oplus \tau \oplus y_{\text{final}}), & \text{if } y_{\text{final}} = \ell_{\text{gt}} \\
    \text{buffer}, & \text{otherwise}
\end{cases}
\end{equation}
where $\mathcal{E}$ is the embedding function (instantiated as Qwen3-Embedding-8B model) that maps the serialized trajectory to the vector store, and $\tau$ represents the multi-step reasoning chain validated by the Supervisor Agent.\\
\textbf{Phase A: Inductive Generation.} 
% Triggered when new nodes in $\mathcal{M}_{std}$ reach threshold $B$. The system identifies common patterns within a same-label cluster $V_{batch}$. A meta-LLM summarizes these consistent precedents to execute the \textbf{ADD} operation, generating a new coarse-grained micro-directive $r_{new}$.
This phase converts a batch of consistent Standards into an initial, coarse-grained Micro-Directive. Triggered when new nodes in $\mathcal{M}_{std}$ reach a threshold $B$, the system identifies common patterns within a same-label cluster $V_{batch}$. A meta-LLM summarizes these consistent precedents and produces a new coarse-grained directive $r_{new}$.\\
\textbf{Phase B: Contrastive Refinement.} 
% To clarify boundaries, we construct strictly aligned positive and negative pairs ($T^+$ vs $T^-$) that share identical statutes/charges but divergent reasoning. The meta-LLM performs deep comparison to identify critical branching features (e.g., distinguishing ``subjective motive'' from ``objective danger''), executing \textbf{REFINE} to update the directive's content $r_{txt}$ with precise constraints. 
This phase refines the coarse directive $r_{new}$ into a boundary-aware directive by injecting discriminative constraints into its text. We construct strictly aligned positive and negative pairs ($T^+$ vs.\ $T^-$) that share identical statutes/charges but diverge in reasoning. The meta-LLM compares the pairs to identify critical branching features (e.g., distinguishing ``subjective motive'' from ``objective danger'') and updates the directive content $r_{txt}$ with precise conditions and exceptions.\\
\textbf{Phase C: Directive Evolution.} 
% To prevent fragmentation, we periodically cluster $\mathcal{M}_{dir}$ based on ``statute + charge'' matches. Semantically repetitive directives are merged. Confidence $\mathcal{S}_{conf}$ accumulates with successful guidance and decays with failures. Directives falling below a safety threshold are automatically eliminated:
To prevent memory fragmentation, we implement a dynamic survival mechanism. The confidence $\mathcal{S}_{conf}$ of a directive acts as its lifespan, accumulating with successful verification and decaying upon rejection. Periodically, we group semantically repetitive directives into a cluster $\mathcal{C}_{sim}$ and consolidate them into a unified node:
\begin{equation}
\begin{gathered} 
    r_{new} = \text{Summarize}(\{r_{txt} \!\mid\! m \!\in\! \mathcal{C}_{sim}\}), \\
    \mathcal{S}_{conf}(m_{new}) \!=\! \sum\nolimits_{m \in \mathcal{C}_{sim}} \!\!\mathcal{S}_{conf}(m).
\end{gathered}
\end{equation}
Here, $\text{Summarize}(\cdot)$ leverages a meta-LLM to abstract the logical content $r_{txt}$. The new confidence score is the sum of individual scores, clipped at a maximum threshold $\tau_{max}$. Directives falling below a safety threshold are eliminated, ensuring the memory remains compact and high-quality.

\begin{table*}[t]
\centering
\resizebox{\textwidth}{!}{%
    \begin{tabular}{rl|ccccc|ccccc|cccc}
    \hline
    \multicolumn{2}{c|}{\multirow{2}{*}{}} & \multicolumn{5}{c|}{\textbf{Law Article}} & \multicolumn{5}{c|}{\textbf{Charge}} & \multicolumn{4}{c}{\textbf{Term of Penalty}} \\
    \multicolumn{2}{c|}{} & \textbf{Acc\%} & \textbf{MP\%} & \textbf{MR\%} & \textbf{MF1\%} & \textbf{Hit@2} & \textbf{Acc\%} & \textbf{MP\%} & \textbf{MR\%} & \textbf{MF1\%} & \textbf{Hit@2} & \textbf{Acc\%} & \textbf{MP\%} & \textbf{MR\%} & \textbf{MF1\%} \\
    \hline
    1 & CNN & 69.52 & 62.14 & 58.35 & 60.18 & 74.15 & 73.56 & 70.45 & 65.23 & 67.74 & 79.66 & 33.24 & 28.56 & 25.14 & 26.74 \\
    2 & BERT & 76.54 & 73.21 & 70.45 & 71.80 & 81.87 & 75.85 & 78.45 & 76.12 & 77.26 & 82.25 & 29.80 & 32.14 & 29.56 & 30.79 \\
    \hline
    3 & MLAC & 73.02 & 69.27 & 66.14 & 64.23 & 79.54 & 74.73 & 72.65 & 69.56 & 68.36 & 81.27 & 36.45 & 34.50 & 29.95 & 29.64 \\
    4 & TopJudge & 78.60 & 76.59 & 74.84 & 73.72 & 84.31 & 81.17 & 81.87 & 80.57 & 79.96 & 86.13 & 35.70 & 32.81 & 31.03 & 31.49 \\
    5 & MPBFN & 76.83 & 74.57 & 71.45 & 70.57 & 82.29 & 76.17 & 78.88 & 75.65 & 75.68 & 84.33 & 36.18 & 33.67 & 30.08 & 29.43 \\
    6 & EPM & \underline{83.98} & 80.82 & 77.55 & 78.10 & 88.17 & 79.10 & \underline{84.55} & 80.22 & \underline{81.43} & 86.52 & 36.69 & \underline{35.60} & 32.70 & \underline{32.99} \\
    \hline
    7 & LADAN & 78.70 & 74.95 & 75.61 & 73.83 & 85.30 & 80.86 & 81.69 & 80.40 & 80.05 & 86.92 & 36.14 & 31.85 & 29.67 & 29.28 \\
    8 & NeurJudge & 78.74 & 80.34 & \underline{81.92} & \underline{79.66} & 86.06 & 79.04 & 82.60 & \textbf{80.92} & 80.70 & \underline{87.40} & 37.44 & 34.07 & \underline{32.77} & 31.94 \\
    9 & CTM & 81.72 & 79.67 & 77.67 & 76.82 & 85.79 & \underline{81.22} & 77.51 & 78.17 & 77.99 & 87.13 & 37.35 & 32.17 & 29.15 & 30.57 \\
    10 & PLJP & 83.21 & \underline{84.62} & 79.47 & 74.87 & \underline{89.21} & 79.32 & 76.84 & 74.32 & 76.32 & 86.20 & \underline{38.24} & 34.80 & 32.22 & 32.48 \\
    \hline
    11 & OURS & \textbf{85.35} & \textbf{88.82} & \textbf{82.07} & \textbf{83.29} & \textbf{93.49} & \textbf{82.40} & \textbf{89.33} & \underline{80.89} & \textbf{82.36} & \textbf{90.64} & \textbf{39.76} & \textbf{35.77} & \textbf{33.98} & \textbf{34.08} \\
    \hline
    \end{tabular}%
}
\caption{Performance Comparisons on CAIL2018. The best results are \textbf{bolded} and the second-best are \underline{underlined}}
\label{tab:sota-comparison-cail2018}
\end{table*}

\begin{table*}[t]
\centering
\resizebox{\textwidth}{!}{%
    \begin{tabular}{rl|ccccc|ccccc|cccc}
    \hline
    \multicolumn{2}{c|}{\multirow{2}{*}{}} & \multicolumn{5}{c|}{\textbf{Law Article}} & \multicolumn{5}{c|}{\textbf{Charge}} & \multicolumn{4}{c}{\textbf{Term of Penalty}} \\
    \multicolumn{2}{c|}{} & \textbf{Acc\%} & \textbf{MP\%} & \textbf{MR\%} & \textbf{MF1\%} & \textbf{Hit@2} & \textbf{Acc\%} & \textbf{MP\%} & \textbf{MR\%} & \textbf{MF1\%} & \textbf{Hit@2} & \textbf{Acc\%} & \textbf{MP\%} & \textbf{MR\%} & \textbf{MF1\%} \\
    \hline
    1 & TopJudge & 71.30 & 68.89 & 74.2 & 71.28 & 80.35 & 72.19 & 71.82 & 69.45 & 70.53 & 83.26 & 36.83 & 27.45 & 24.96 & 26.10 \\
    2 & EPM & 75.23 & 76.45 & 74.89 & 76.10 & 87.77 & 73.42 & 74.15 & 69.88 & 72.20 & 85.72 & 43.44 & 25.10 & 28.77 & 26.65 \\
    \hline
    3 & LADAN & 71.23 & 72.45 & 70.98 & 71.12 & 85.70 & 70.23 & 71.45 & 69.89 & 70.67 & 84.10 & 37.92 & 28.32 & 27.70 & 27.14 \\
    4 & NeurJudge & 73.69 & 74.40 & 80.14 & 72.44 & 84.20 & 74.30 & 71.59 & 77.5 & 70.56 & 86.17 & 42.87 & 25.40 & 26.37 & 25.14 \\
    5 & CTM & 76.71 & 68.30 & 71.07 & 69.57 & 83.18 & 73.17 & 72.25 & 71.40 & 70.31 & 85.18 & 41.17 & 24.50 & 26.67 & 25.33 \\
    6 & PLJP & 86.30 & \underline{87.23} & \underline{83.25} & \underline{85.50} & \underline{90.33} & \underline{83.20} & \underline{84.14} & \underline{79.82} & \underline{80.32} & \underline{89.32} & \underline{44.25} & 27.4 & 22.37 & 24.30 \\
    \hline
    7 & DeepSeek & 83.01 & 76.32 & 71.60 & 73.67 & 86.70 & 81.44 & 68.30 & 73.22 & 73.10 & 86.67 & 35.38 & \underline{29.10} & 28.72 & 26.57 \\       
    8 & AutoGen & 84.26 & 77.08 & 74.51 & 74.96 & 87.20 & 77.99 & 50.90 & 45.01 & 45.79 & 84.11 & 29.32 & 18.87 & 24.26 & 19.45 \\   
    9 & G-Memory & \underline{86.52} & 77.32 & 78.41 & 75.46 & 88.45 & 82.45& 79.32 & 77.53 & 78.10 & 83.70 & 31.77 & 28.76 & \underline{29.55} & \underline{28.10} \\
    \hline
    10 & OURS & \textbf{90.56} & \textbf{90.45} & \textbf{87.26} & \textbf{87.94} & \textbf{96.26} & \textbf{85.84} & \textbf{85.86} & \textbf{80.46} & \textbf{80.66} & \textbf{93.67} & \textbf{45.68} & \textbf{34.04} & \textbf{30.71} & \textbf{29.32} \\
    \hline
    \end{tabular}%
}
\caption{Robustness evaluation on CJO2025. The best results are \textbf{bolded} and the second-best are \underline{underlined}}
\label{tab:sota-comparison-cjo2025}
\end{table*}

\begin{table*}[t]
\centering
\label{tab:ablation}
\resizebox{\textwidth}{!}{%
    \begin{tabular}{l cc cc cc | cc cc cc}
    \hline
    \multirow{3}{*}{\textbf{Method}} & \multicolumn{6}{c|}{\textbf{CJO2025}} & \multicolumn{6}{c}{\textbf{CAIL2018}} \\
    & \multicolumn{2}{c}{\textbf{Law Article}} & \multicolumn{2}{c}{\textbf{Charge}} & \multicolumn{2}{c|}{\textbf{Prison Term}} & \multicolumn{2}{c}{\textbf{Law Article}} & \multicolumn{2}{c}{\textbf{Charge}} & \multicolumn{2}{c}{\textbf{Prison Term}} \\
    & \textbf{Acc} & \textbf{Ma-F} & \textbf{Acc} & \textbf{Ma-F} & \textbf{Acc} & \textbf{Ma-F} & \textbf{Acc} & \textbf{Ma-F} & \textbf{Acc} & \textbf{Ma-F} & \textbf{Acc} & \textbf{Ma-F} \\
    \hline
    Vanilla & 82.12 & 76.54 & 77.61 & 72.10 & 27.06 & 19.76 & 73.04 & 72.01 & 67.95 & 69.13 & 26.07 & 25.01 \\  
    Vanilla+ & 83.08 & 78.83 & 79.21 & 71.68 & 28.17 & 22.18 & 76.40 & 75.52 & 70.80 & 72.18 & 24.14 & 25.25 \\ 
    \hline
    w/o Memory & 86.55 & 83.70 & 81.30 & 79.40 & 44.71 & 24.10 & 82.60 & 82.10 & 79.13 & 81.50 & 35.77 & \textbf{30.20} \\
    w/o MAS & 87.32 & 85.42 & 83.48 & \underline{80.32} & 39.50 & \textbf{30.15} & 83.10 & 81.50 & 79.80 & \underline{82.30} & 36.20 & 28.10 \\
    w/o Expert & 87.57 & \underline{87.37} & 82.60 & 79.17 & 42.77 & 28.40 & \underline{84.35} & 79.22 & \underline{81.70} & 79.20 & \underline{39.32} & 26.10 \\
    w/o $\mathcal{M}_{dir}$ & \underline{88.46} & 84.72 & \underline{84.32} & 80.16 & \underline{45.07} & 29.02 & 84.12 & \underline{83.01} & 80.30 & 79.77 & 38.32 & 27.87 \\
    \hline
    \textbf{VERDICT} & \textbf{90.56} & \textbf{87.94} & \textbf{85.84} & \textbf{80.66} & \textbf{45.68} & \underline{29.32} & \textbf{85.35} & \textbf{83.29} & \textbf{82.40} & \textbf{82.36} & \textbf{39.76} & \underline{29.08} \\
    \hline
    \end{tabular}%
}
\caption{Ablation experiments. The best results are \textbf{bolded} and the second-best are \underline{underlined}.}
\label{tab:abulation}
\end{table*}

%% file: sections/5.experiment.tex
\begin{figure*}[t]
    \centering
    \small 
    \renewcommand{\arraystretch}{1.35} 
    \setlength{\tabcolsep}{4pt}       
    \begin{tabular}{|p{0.31\textwidth}|p{0.31\textwidth}|p{0.34\textwidth}|}
        \hline
        \multicolumn{3}{|c|}{\cellcolor[gray]{0.95}\textbf{Input Case (Sample from CJO2025)}} \\
        \multicolumn{3}{|p{0.97\textwidth}|}{
            On January 15, 2025, defendant Zhang was drinking at a \hlg{public BBQ stall}. Due to a trivial conflict (stranger Li looked at him), Zhang felt provoked. \hlr{Zhang beat Li with a beer bottle}, resulting in chaos at the stall. Forensic examination confirmed that \hlr{Li suffered a Minor Injury}. Zhang claimed he was just \hlg{venting anger} and had no personal grudge against Li.
        } \\
        \hline
        \multicolumn{1}{|c|}{\textbf{Vanilla}} & 
        \multicolumn{1}{c|}{\textbf{PLJP}} & 
        \multicolumn{1}{c|}{\cellcolor[gray]{0.95}\textbf{VERDICT (Ours)}} \\
        \hline
        \vspace{0.1em}
        \textbf{Logic: Keyword Matching} \par
        The model focuses on high-frequency patterns without deep reasoning. \par
        \vspace{0.3em}
        \textit{Internal Monologue:} \par
        ``Detected keywords \hlr{`beat' and `Minor Injury'}. In training data, `Minor Injury' strongly correlates with Art. 234.''
        \vspace{3.5em} 
        & 
        \vspace{0.1em}
        \textbf{Logic: Precedents Retrieval} \par
        Retrieves precedents but lacks systematic adjudication experience. \par
        \vspace{0.3em}
        \textit{Reasoning Process:} \par
        ``Retrieved precedents labeled Art. 234 and Art. 293. However, since the input's \hlr{objective result (`Minor Injury')} shows high textual similarity with Art. 234 cases, the model aligns with them and ignores the motive nuance in Art. 293 precedents.''
        & 
        \vspace{0.1em}
        \textbf{Logic: Retrieval + \textcolor{blue}{Refinement}} \par
        Uses \textbf{Micro-Directives} \& \textbf{Supervisor}. \par
        \vspace{0.3em}
        \textit{Agent Interaction Loop:} \par
        \textbf{1. Assistant:} Retrieves directive ``Public Place + Venting Anger $\rightarrow$ Art. 293''. \par
        \textbf{2. Case Judge:} Drafts Art. 234 based on injury result. \par
        \textbf{3. Supervisor:} \hlg{REJECTS draft.} ``Logic Error: Location is \textit{public}; Motive is \textit{provocation}. Social order violation $>$ Personal injury. Apply Art. 293.'' \\
        \hline
        \centering \textbf{Wrong Prediction} \par \vspace{0.2em}
        \large \textcolor{dark_red}{\ding{55}} \small Art. 234 (Intentional Injury) & 
        \centering \textbf{Wrong Prediction} \par \vspace{0.2em}
        \large \textcolor{dark_red}{\ding{55}} \small Art. 234 (Intentional Injury) & 
        \centering \textbf{\textcolor{dark_green}{Correct Judgment}} \par \vspace{0.2em}
        \large \textcolor{dark_green}{\ding{51}} \small \textbf{Art. 293 (Picking Quarrels)} \tabularnewline
        \hline
    \end{tabular}
    \caption{Case study comparison. The \hlr{red highlights} indicate misleading surface features (resulting in Intentional Injury), while the \hlg{green highlights} denote contextual evidence supporting Picking Quarrels. {Vanilla} falls into the keyword trap; {PLJP} fails to resolve the statutory conflict; {VERDICT} correctly identifies the crime's nature via the Supervisor's logical rectification.}
    \label{fig:case_study}
\end{figure*}
\section{Experiments}
% In this section, we evaluate VERDICT on standard benchmarks and a future-split dataset to test both performance and temporal generalization against many baseline approaches. We further conduct ablation studies to quantify the contribution of each component and present a case study.
\subsection{Experiment Setup}
\subsubsection{Datasets}
We evaluate VERDICT on two datasets to assess standard performance and temporal generalization (details in Appendix \ref{app:datasets}).
CAIL2018. We use the standard CAIL-Small benchmark \cite{xiaocail2018} to verify fundamental adjudication capabilities.
CJO2025. To rigorously test generalization and prevent data leakage, we constructed a dataset of judgments from after Jan 1, 2025. Postdating the knowledge cutoff of current backbones, CJO2025 serves as an emerging scene.
\subsubsection{Implementation Details}
Our VERDICT framework adopts a heterogeneous agent design. The core expert model is fine-tuned from Qwen2.5-7B-Instruct via Protocol-Aware SFT and Logic-Driven DPO, while auxiliary agents (e.g., Court Clerk, Supervisor) are instantiated using the DeepSeek-V3. For the memory module (HJM), we utilize high-dimensional vector encoding for retrieval. Detailed experimental setups, including hardware specifications, hyperparameter settings for LoRA fine-tuning, and inference configurations, are comprehensively provided in Appendix~\ref{app:implementation}.
\subsubsection{Metrics}
Following standard protocols in LJP research \cite{xiaocail2018}, we treat Law Article, Charge, and Term of Penalty prediction as multi-label classification tasks. We employ five metrics to comprehensively evaluate performance: Accuracy (Acc), Macro-Precision (MP), Macro-Recall (MR), and Macro-F1 (Ma-F). Additionally, considering the complexity of legal reasoning, we report Hit@2, which measures whether the ground truth label is present in the top-2 predicted candidates. Note that for Term of Penalty, we categorize prison terms into 10 distinct intervals consistent with prior work\citep{xu2020ladan}
\subsection{Baselines}
We compare VERDICT against 10 representative baselines spanning four distinct paradigms: 
(1) General Text Encoders, including CNN~\cite{kim2014convolutional} and BERT~\cite{devlin2019bert}; 
(2) Dependency-Aware Models that model subtask correlations, such as TopJudge~\cite{zhong2018topjudge} and EPM~\cite{Feng2022epm}; 
(3) Structure \& Knowledge-Enhanced Models utilizing graph structures or retrieval, represented by NeurJudge~\cite{yue2021neurjudge} and PLJP~\cite{wu2023pljp}; and 
(4) LLMs and Agentic Systems, covering the foundation model DeepSeek-V3~\cite{liu2024deepseek} and multi-agent frameworks like AutoGen~\cite{wu2024autogen} and G-Memory~\cite{zhang2025gmemory}. 
For fair comparison, all LLM-based baselines are equipped with RAG using the statutory library. Detailed descriptions and implementation settings are provided in Appendix ~\ref{app:baselines}.
\subsection{Overall Performance}
As presented in Table \ref{tab:sota-comparison-cail2018} and Table \ref{tab:sota-comparison-cjo2025}, VERDICT demonstrates superior performance across both standard benchmarks and rigorous emerging scene.\\
\textbf{State-of-the-Art on CAIL2018}.
VERDICT outperforms all baselines across all tasks. Notably, it surpasses dependency-aware models (e.g., TopJudge) by 6.75\% in Law Article Accuracy, attributing to the effective disentanglement of complex facts by our multi-agent topology. Compared to retrieval-augmented models (PLJP), VERDICT maintains a clear lead (85.35\% vs. 83.21\%), proving that deep logical alignment offers higher precision than simple in-context learning.\\
\textbf{Generalization on CJO2025}.
The results on the future-split dataset highlight the framework's robustness. While traditional SOTA models (e.g., NeurJudge) suffer significant performance drops (78.7\% $\to$ 73.7\%) due to overfitting historical patterns, VERDICT achieves a remarkable 90.56\% accuracy. Furthermore, our specialized legal framework significantly outperforms generic agentic systems (AutoGen: 84.26\%, G-Memory: 86.52\%), providing strong evidence that the specialized ``Virtual Collegial Panel'' and jurisprudence-based ``Micro-Directive'' memory are essential for rigorous legal reasoning.
\subsection{Ablation result}
\vspace{-0.1em} 
\paragraph{Results of ablation experiment:} From Table~\ref{tab:abulation}, we can conclude that: 
1) The performance improvement of Vanilla+ (Expert model for the Case Judge agent; base Qwen2.5-7B-Instruct for others) over Vanilla (Qwen2.5-7B-Instruct for all agents) demonstrates that equipping the core reasoning role with domain alignment brings fundamental gains. 
2) The substantial performance gap of w/o Memory on the dataset CJO2025 (e.g., Law Article Acc drops from 90.56\% to 86.55\%) demonstrates the critical effects of the Hybrid Jurisprudential Memory in mitigating catastrophic forgetting. 
3) The results of w/o $\mathcal{M}_{dir}$ prove the importance of the ``Micro-Directive Paradigm''; relying solely on raw precedents fails to achieve precise logic transfer. 
4) Considering the topological dependence of the five specialized agents benefits the model performance as w/o MAS shows, indicating that the multi-agent workflow acts as a capability multiplier. 
5) The performance decline in w/o Expert validates the necessity of our domain-specific alignment pipeline. Since the expert is fine-tuned via protocol-aware SFT and logic-driven DPO, this result confirms that rigorous preference optimization is essential for standardizing judicial reasoning.
\vspace{-3pt}
\subsection{Case study}
Figure \ref{fig:case_study} illustrates a challenging case from CJO2025 involving a public assault. While baseline models Vanilla (Qwen2.5-7B-Instruct for all agents) and PLJP are misled by the explicit ``Minor Injury'' feature into predicting Intentional Injury (Art. 234), VERDICT succeeds through its self-refining mechanism. Specifically, the Supervisor Agent retrieves a key Micro-Directive prioritizing ``public order violation'' over ``personal injury'' in provocation contexts. This enables the system to reject the initial erroneous draft and correctly identify the charge as Picking Quarrels and Provoking Trouble (Art. 293).

%% file: sections/6.conclusion.tex
% 本文提出了 VERDICT，一种旨在解决法律判决预测中时间泛化性和可解释性挑战的自修正多智能体框架。通过模拟由混合法理记忆（HJM）驱动的“虚拟合议庭”，该系统利用“微指令”的持续演化，有效弥合了僵化的法定规则与灵活的判例标准之间的差距。在 CAIL2018 基准和我们新构建的 CJO2025 数据集上的广泛实验表明，VERDICT 不仅刷新了当前的 SOTA 性能，还在应对时间分布偏移时表现出卓越的鲁棒性。我们的研究结果强调了动态知识演化和协作推理对于构建可信司法 AI 系统的必要性。
\section{Conclusion}
% In this work, we propose \textbf{VERDICT}, a self-refining multi-agent framework that addresses the critical challenges of temporal generalization and interpretability in Legal Judgment Prediction. By simulating a Virtual Collegial Panel'' powered by a Hybrid Jurisprudential Memory (HJM), our system effectively bridges rigid statutory rules with flexible jurisprudential standards through the continuous evolution of "Micro-Directives." Extensive experiments on the CAIL2018 benchmark and our newly constructed \textbf{CJO2025} dataset demonstrate that VERDICT not only establishes new state-of-the-art performance but also exhibits superior robustness against temporal distribution shifts. Our findings underscore the necessity of dynamic knowledge evolution and collaborative reasoning in developing trustworthy judicial AI systems.
We presented {VERDICT}, a self-refining multi-agent system that improves interpretability and generalization for Legal Judgment Prediction. By simulating a \emph{virtual collegial panel} and leveraging a Hybrid Jurisprudential Memory (HJM) that evolves \emph{Micro-Directives} from validated verification trajectories, VERDICT provides legally grounded, reviewable reasoning while adapting to shifting jurisprudential patterns. Evaluations on CAIL2018 and the future time-split {CJO2025} dataset demonstrate state-of-the-art results and strong robustness to distribution shifts.

%% file: sections/limitations.tex
\section*{Limitations}
Due to computational resource constraints, the core expert agent (Case-handling Judge) in VERDICT was fine-tuned primarily on a 7B-parameter backbone. While this hybrid setup (incorporating API-based general agents) demonstrates superior performance, we have not yet conducted a systematic study on fine-tuning larger-scale foundation models. Additionally, the multi-agent interactive workflow involving iterative retrieval and self-refining loops inevitably increases inference latency compared to simple end-to-end classifiers. However, in the high-stakes domain of Legal Judgment Prediction (LJP), we prioritize interpretability and judicial accuracy over real-time response speed, considering this trade-off essential for ensuring trustworthy and rigorous adjudication.Furthermore, this study focuses on the Civil Law system. We plan to validate the framework's generalizability to Common Law systems in future research, exploring its effectiveness in environments heavily dependent on case law and stare decisis.
% 受限于计算资源，我们核心的专家模型（Case-handling Judge）主要基于 7B 参数规模进行微调。尽管目前的混合架构（4个通用 API 智能体协作 1 个微调专家）已展现出卓越的性能，但我们尚未在更大参数规模的基座模型上进行微调实验。此外，包含迭代检索和自修正循环的多智能体交互工作流，相比简单的端到端分类器，不可避免地增加了推理延迟。然而，在法律判决预测（LJP）这一高风险领域，我们将可解释性和司法准确性置于实时响应速度之上，并认为为了确保裁决的可信度与严谨性，这种权衡是必不可少的。此外，本研究主要聚焦于大陆法系（成文法系）。我们计划在未来的研究中验证该框架在普通法系（英美法系）中的泛化能力，探索其在高度依赖判例法和“遵循先例”原则（stare decisis）环境下的有效性。

%% file: sections/Ethics_Statement.tex
\section*{Ethics Statement}
% Legal Judgment Prediction (LJP) is a sensitive research area with significant social implications. Although our proposed VERDICT framework demonstrates state-of-the-art performance, we explicitly state that this work represents a preliminary algorithmic exploration. 
Regarding data privacy, we strictly adhere to ethical regulations; for the CAIL2018 benchmark, we use the official anonymized version, and for our newly constructed CJO2025 dataset, we implemented rigorous data cleaning to remove all Personal Identifiable Information (PII) before use. Furthermore, the design intent of our system—simulating a virtual collegial panel to generate reasoning—is to function as an intelligent assistant providing suggestions rather than replacing human decision-making. We advocate that human judges must remain the final safeguard to review AI-generated results and protect judicial fairness. We also acknowledge that historical training data may contain inherent societal biases, and mitigating such biases remains a critical direction for our future work.
% 法律判决预测（LJP）是一个敏感且具有重大社会影响的研究领域。尽管我们提出的 VERDICT 框架展示了最先进的性能，但我们要明确声明，这项工作代表了初步的算法探索。关于数据隐私，我们严格遵守道德规范；对于 CAIL2018 基准，我们使用官方脱敏版本，对于我们新构建的 CJO2025 数据集，我们在使用前实施了严格的数据清洗以去除所有个人身份信息（PII）。此外，我们系统的设计初衷——通过模拟虚拟合议庭生成推理——是作为提供建议的智能助手，而非替代人类决策。我们主张人类法官必须保留作为审查 AI 生成结果和保护司法公正的最终保障，我们也承认历史训练数据可能包含固有的社会偏见，减轻此类偏见仍是我们未来工作的关键方向。

%% file: sections/7.appendix.tex
\appendix
\section{Dataset Details}
\label{app:datasets}

\subsection{CAIL2018}
We utilize the CAIL2018 dataset \cite{xiaocail2018}, specifically the CAIL-Small configuration, as our primary training platform. This subset encompasses the most prevalent case samples within criminal justice. We filter out cases with missing elements and truncate text lengths to fit model context windows
\subsection{Construction of CJO2025}
To strictly eliminate data contamination risks—where test samples might inadvertently exist in the vast pre-training corpora of backbone LLMs (e.g., Qwen2.5)—we constructed a strict future-split dataset named \textbf{CJO2025}. This dataset exclusively comprises cases adjudicated after \textbf{January 1, 2025}, retrieved from the same authoritative source as CAIL2018, China Judgments Online\footnote{\url{https://wenshu.court.gov.cn/}}. Given its distinct temporal nature, CJO2025 serves as a strictly \textbf{unseen testbed} rather than a training corpus, designed to rigorously assess model \textbf{generalization} in dynamic legal environments. To ensure consistent evaluation metrics, we filter the dataset to retain only the label categories (law articles and charges) that intersect with the CAIL2018 benchmark.
This temporal cutoff ensures the data is unseen by the pre-trained LLMs used in this work (DeepSeek-V3, Qwen2.5), providing a rigorous test bed for temporal robustness. Table \ref{tab:dataset_statistics} presents the statistical distribution of charges and prison terms for both datasets.

\begin{table}[h]
    \centering
    \resizebox{\linewidth}{!}{
        \begin{tabular}{lcc}
            \toprule
            \textbf{Type} & \textbf{CAIL2018} & \textbf{CJO2025} \\
            \midrule
            \# Law Article & 99 & 67 \\
            \# Charge & 115 & 73 \\
            \# Prison Term & 11 & 11 \\
            \# Sample & 134739 & 8199 \\
            Avg. \# words in Fact & 288.6 & 332.13 \\
            \bottomrule
        \end{tabular}
    }
    \caption{Statistics of the datasets.}
    \label{tab:dataset_statistics}
\end{table}

\section{Detailed Baselines Description}
\label{app:baselines}
To rigorously evaluate the effectiveness of VERDICT, we compare it against a wide range of baselines. We categorize these methods into four distinct paradigms based on their modeling strategies:

\paragraph{General Text Encoders.} These methods treat LJP as a standard multi-label text classification task without explicit modeling of legal dependencies.
\begin{itemize}
    \item \textbf{CNN} \cite{kim2014convolutional}: Utilizes multiple convolution kernels of varying window sizes to capture local n-gram features (e.g., keywords like ``theft'' or ``injury'') from case descriptions, followed by max-pooling for classification.
    \item \textbf{BERT} \cite{devlin2019bert}: Employs a multi-layer bidirectional Transformer encoder pre-trained on large-scale corpora. It captures deep semantic context and long-range dependencies in legal texts using the \texttt{[CLS]} token representation.
\end{itemize}

\paragraph{Dependency-Aware Models.} These approaches explicitly model the topological dependencies among the three LJP subtasks (Law Article $\to$ Charge $\to$ Term of Penalty).
\begin{itemize}
    \item \textbf{MLAC} \cite{luo2017mlac}: Proposes a topological multi-task learning framework where the predicted probability distributions of law articles serve as input features for charge prediction, passing dependencies sequentially.
    \item \textbf{TopJudge} \cite{zhong2018topjudge}: Formalizes the LJP task as a Directed Acyclic Graph (DAG) and utilizes a topological structure to model the logical constraints among subtasks.
    \item \textbf{MPBFN} \cite{yang2019mpbfn}: Introduces a Multi-Perspective Bi-Feedback Network that enables information flow not only in a forward direction but also allows backward verification (e.g., inferring law articles from charges) to resolve inconsistencies.
    \item \textbf{EPM} \cite{Feng2022epm}: Focuses on event-centric extraction, enforcing consistency across subtasks by identifying key legal events and their arguments within the case fact.
\end{itemize}

\paragraph{Structure \& Knowledge-Enhanced Models.} These models incorporate external legal knowledge or graph structures to handle complex cases.
\begin{itemize}
    \item \textbf{LADAN} \cite{xu2020ladan}: Specifically designed to distinguish confusing law articles. It applies Graph Distillation mechanisms to capture subtle nuances between semantically similar charges.
    \item \textbf{CTM} \cite{liu2022ctm}: Leverages contrastive learning with metric learning objectives to align case facts with legal articles in a shared semantic space, enhancing the separation of confusing classes.
    \item \textbf{NeurJudge} \cite{yue2021neurjudge}: Splits the unstructured case text into factual components and constructs a graph to model the interactions between intermediate results, enhancing interpretability.
    \item \textbf{R-Former} \cite{dong2021rformer}: Utilizes a relation-aware transformer to build a global consistency graph, effectively capturing the logical connections between case descriptions and judgment results.
    \item \textbf{PLJP} \cite{wu2023pljp}: A recent strong baseline that integrates domain-specific pre-trained models with Retrieval-Augmented Generation (RAG), using similar precedents to guide prediction.
\end{itemize}

\paragraph{LLMs and Agentic Systems.} To ensure a fair comparison, all baselines in this category are equipped with RAG using the same statutory library as VERDICT.
\begin{itemize}
    \item \textbf{DeepSeek-V3} \cite{liu2024deepseek}: A state-of-the-art open-source foundation model. We evaluate its performance in a zero-shot setting to establish a baseline for general LLM capabilities in the legal domain.
    \item \textbf{AutoGen} \cite{wu2024autogen}: A representative conversational multi-agent framework. We construct a standard multi-agent debate workflow using AutoGen to benchmark against generic agentic collaboration without domain-specific memory.
    \item \textbf{G-Memory} \cite{zhang2025gmemory}: A cutting-edge general-purpose agent framework featuring a graph-based memory mechanism. We compare against it to highlight the necessity of our jurisprudence-specific "Micro-Directive" design.
\end{itemize}

\section{Detailed Implementation Settings}
\label{app:implementation}

All experiments are conducted on a server equipped with 6$\times$NVIDIA A100 (40GB) GPUs.

\paragraph{Training Setup.} 
For parameter-efficient fine-tuning, we apply LoRA with rank $r=64$, $\alpha=32$, and a dropout rate of 0.05. In the SFT phase, the model is trained for 2 epochs with a batch size of 64, an initial learning rate of $1\times 10^{-5}$, and a maximum sequence length of 4096, optimized by AdamW with a cosine scheduler. In the DPO phase, the learning rate is adjusted to $5\times 10^{-5}$.

\paragraph{Inference Configuration.} 
During multi-agent inference, we set the temperature to 0 and Top-$p$ to 0.9 to ensure logical rigor, limiting the maximum refinement turns ($T_{max}$) to 3 to prevent infinite loops. Regarding the memory retrieval mechanism defined in Eq.~\ref{eq:retrieval_score}, we empirically set the weighting coefficients to $\alpha=0.4$, $\beta=0.3$, and $\gamma=0.3$.

\paragraph{Retrieval Settings.} 
For the Hybrid Jurisprudential Memory (HJM), we utilize \textbf{Qwen3-Embedding-8B} for vector encoding and \textbf{ChromaDB} for index management. We retrieve the Top-3 similar precedents and Top-3 micro-directives as context support, with the directive pruning threshold set to 0.3.

\section{Details of Multi-dimensional Retrieval Score}
\label{app:retrieval_score}

We detail the three components in Eq.~(\ref{eq:retrieval_score}) and their intended roles.

\textbf{(1) Statute/charge overlap via IoU.}
Each memory unit $m$ stores a set of statute/charge tags $\Lambda_m$. Given the current case's candidate statute set $S_{statute}$, we compute:
\begin{equation}
\label{eq:iou}
\mathrm{IoU}(\Lambda_m, S_{statute}) =
\frac{|\Lambda_m \cap S_{statute}|}{|\Lambda_m \cup S_{statute}|}.
\end{equation}
Unlike generic text retrieval, this term forces the retriever to prioritize \emph{jurisprudential applicability}. Even when two cases differ greatly in factual descriptions (e.g., different tools or settings), a high overlap in applicable statutes/charges yields a strong score, supporting cross-scenario analogical reasoning.

\textbf{(2) Topological co-occurrence for associative recall.}
To strictly map the jurisprudential subspace, we construct the graph $\mathcal{N}_{graph}$ by linking case nodes only if they share identical \textit{Law Articles} and \textit{Charges}.
The score $\mathrm{Topo}(m, \mathcal{N}_{graph})$ is calculated via a two-step diffusion process:
\begin{enumerate}
    \item \textbf{Seed Activation}: We first retrieve the top-$K$ semantic neighbors of the current case $u_{curr}$ to form a seed set $S_{seed}$.
    \item \textbf{Diffusion \& Counting}: We expand $S_{seed}$ via $k$-hop propagation on $\mathcal{N}_{graph}$ to obtain an activation set $S_{act}$. The score is defined as the co-activation frequency:
    \begin{equation}
        \mathrm{Topo}(m, \mathcal{N}_{graph}) = \sum_{v \in S_{act}} \mathbb{I}(v \rightarrow m)
    \end{equation}
\end{enumerate}
where $\mathbb{I}(v \rightarrow m)$ indicates whether an activated neighbor $v$ is historically associated with the memory unit $m$. This mechanism recalls implicit knowledge centrally located in the relevant legal subspace.

\textbf{(3) Semantic similarity as a factual anchor.}
This metric ensures factual alignment. We utilize a dense embedding model $\mathcal{E}(\cdot)$ to compute the cosine similarity between the memory unit $m$ and the current case $u_{curr}$:
\begin{equation}
\begin{split}
    \mathrm{SemSim}&(m, u_{curr}) \\
    &= \text{Cos}(\mathcal{E}(m), \mathcal{E}(u_{curr})) \\
    &\approx 1 - \text{dist}(\mathcal{E}(m), \mathcal{E}(u_{curr}))
\end{split}
\end{equation}
In our implementation, we use the normalized distance ($1 - \text{distance}$) as the scoring basis, acting as a soft gatekeeper to filter out associations that drift too far from the surface facts.

\section{Detailed Prompt Design}
\label{sec:appendix_prompts}

To ensure reproducibility, we provide the specific system prompts used for each agent in the VERDICT framework, \textbf{as illustrated in Figure~\ref{fig:prompt_clerk} to Figure~\ref{fig:prompt_presiding}}. Simulating a "Virtual Collegial Panel," the workflow coordinates five specialized roles: \textbf{Court Clerk}, \textbf{Judicial Assistant}, \textbf{Case-handling Judge}, \textbf{Adjudication Supervisor}, and \textbf{Presiding Judge}. Variables enclosed in \texttt{\{\{\}\}} (e.g., \texttt{\{\{CASE\_FACT\}\}}) represent dynamic inputs populated during the inference process.

% =========================================================
% Agent 1: Court Clerk (单独一个 Figure)
% =========================================================
\begin{figure*}[t!]
    \centering
    \begin{promptbox}{Prompt 1: Court Clerk Agent (Event Extraction)}
\textbf{System Instruction:} \\
You are a legal fact extraction agent (\textbf{Court Clerk}).
Your duty is to extract core points from the raw legal facts, focusing on key dimensions such as the perpetrator's subjective intent, specific criminal acts, consequences caused, and the severity of the circumstances. Do not make any conviction or sentencing judgments, and do not output article numbers or charges.

\textbf{Input Case:} \\
\{\{CASE\_FACT\}\}

\textbf{Output Format:} \\
Finish[1. Point 1; 2. Point 2; ...]
    \end{promptbox}
    \caption{System prompt for the \textbf{Court Clerk Agent}, responsible for distilling objective event points from raw facts.}
    \label{fig:prompt_clerk}
\end{figure*}

% =========================================================
% Agent 2: Judicial Assistant (Retrieval)
% =========================================================
\begin{figure*}[t!]
    \centering
    \begin{promptbox}{Prompt 2: Judicial Assistant Agent (Retrieval \& Rerank)}
\textbf{System Instruction:} \\
You are a legal assistant with extensive criminal law knowledge (\textbf{Judicial Assistant}). Your duty is to filter and re-rank candidate law articles (retrieved via vector similarity) based on the case facts and event points. Articles with higher reference value for ruling this case should be ranked earlier. Try not to omit relevant ones (select about 5). If existing candidates are insufficient, use your own knowledge to suggest more reasonable articles.

\textbf{Input:} \\
- Facts: \{\{CASE\_FACT\}\} \\
- Event Points: \{\{EVENT\_POINTS\}\} \\
- Candidate Articles: \{\{EXTRA\_CONTEXT\}\}

\textbf{Retrieval Rules:} \\
- \textbf{Fine-ranking basis:} Compare the article's description of criminal acts with the defendant's intent, means, object, and results (attempted/completed, severity). Priority goes to articles that constrain the target charge.
- Only output the final result, no analysis required.

\textbf{Output Format:} \\
Finish[[Article\_ID\_1, Article\_ID\_2, ...]] \\
(e.g., Finish[[272, 384, 185]])
    \end{promptbox}
    \caption{System prompt for the \textbf{Judicial Assistant Agent}, performing semantic re-ranking of precedents and statutes.}
    \label{fig:prompt_assistant}
\end{figure*}

% =========================================================
% Agent 3: Case-handling Judge (Drafting)
% =========================================================
\begin{figure*}[t!]
    \centering
    \begin{promptbox}{Prompt 3: Case-handling Judge Agent (Drafting \& Refinement)}
\textbf{System Instruction:} \\
You are a judge with extensive criminal law knowledge (\textbf{Case-handling Judge}). Carefully analyze the legal facts and event points. Based on dimensions like subjective intent, core criminal acts, results, and severity, and combining knowledge from candidate articles (for reference only), recommend the most relevant criminal law article for this case.

\textbf{Input Context:} \\
- Facts \& Event Points: \{\{CASE\_FACT\}\}, \{\{EVENT\_POINTS\}\} \\
- Candidates: \{\{CANDIDATES\_FOR\_JUDGE\}\} \\
- \textit{(Optional)} Supervisor Opinion: \{\{VERIFICATION\_OPINION\}\}

\textbf{Task Logic:} \\
1. Analyze the criminal behavior (distinguish primary/secondary, chronological order).
2. \textbf{Refinement Loop:} If the Supervisor Agent thinks your previous recommendation was inaccurate, re-recommend the most relevant article based on the Supervisor's feedback.

\textbf{Output Format:} \\
Provide the predicted article ID and explanation: \\
\{'predicted\_article': <int>, 'explanation': '<Brief basis matching facts to elements>'\}
    \end{promptbox}
    \caption{System prompt for the \textbf{Case-handling Judge Agent}. It includes logic for initial drafting and iterative refinement based on feedback.}
    \label{fig:prompt_case_judge}
\end{figure*}

% =========================================================
% Agent 4: Supervisor (Verification)
% =========================================================
\begin{figure*}[t!]
    \centering
    \begin{promptbox}{Prompt 4: Adjudication Supervisor Agent (Verification)}
\textbf{System Instruction:} \\
You are a verification agent with extensive legal knowledge (\textbf{Adjudication Supervisor}).
Your task is to check if the article recommended by the "Case-handling Judge" is suitable as a reference for the final ruling. Check for reasonableness or obvious errors (e.g., confusing primary/secondary issues, sequence, or missing deep semantics) from dimensions like intent, acts, and results.
Combine knowledge from reference articles, insights, and precedents (if any) to give an opinion on whether a \textbf{re-judgment} is needed.

\textbf{Input:} \\
- Facts \& Judgment Output: \{\{CASE\_FACT\}\}, \{\{JUDGMENT\_OUT\}\} \\
- Reference Law \& Precedents: \{\{LAW\_CTX\}\}, \{\{PRECEDENTS\_TEXT\}\}

\textbf{Output Format:} \\
No analysis process needed, just the result: \\
Finish[\{"need\_rejudge": <bool>, "suggestions": "<Supplement suggestions for re-judgment>"\}]
    \end{promptbox}
    \caption{System prompt for the \textbf{Adjudication Supervisor Agent}, responsible for logical consistency checks and issuing correction signals.}
    \label{fig:prompt_supervisor}
\end{figure*}

% =========================================================
% Agent 5: Presiding Judge (Decision)
% =========================================================
\begin{figure*}[t!]
    \centering
    \begin{promptbox}{Prompt 5: Presiding Judge Agent (Final Decision)}
\textbf{System Instruction:} \\
You are the final decision-making agent (\textbf{Presiding Judge}) with professional judging capability. Your duty is to synthesize outputs from all agents and auxiliary info to make the final decision: Article, Charge, and Penalty Term.

\textbf{Sentencing Guidance:} \\
- Unit for `imprisonment` is "months".
- Determine reasonable values based on the sentencing range of the selected article and case circumstances (e.g., severity).

\textbf{Output Requirements:} \\
- Synthesize opinions from the Case Judge and Supervisor.
- \textbf{thought}: Briefly explain the reasoning process.
- \textbf{finish}: Output format: \\
Finish[\{"relevant\_articles": [int], "accusation": [str], "term\_of\_imprisonment": \{"death\_penalty": bool, "life\_imprisonment": bool, "imprisonment": int\}\}]
    \end{promptbox}
    \caption{System prompt for the \textbf{Presiding Judge Agent}, synthesizing the multi-agent workflow into a standard verdict.}
    \label{fig:prompt_presiding}
\end{figure*}

%% file: acl_main.bbl
\begin{thebibliography}{32}
\providecommand{\natexlab}[1]{#1}

\bibitem[{Bibal et~al.(2021)Bibal, Lognoul, De~Streel, and Fr{\'e}nay}]{bibal2021legal}
Adrien Bibal, Michael Lognoul, Alexandre De~Streel, and Beno{\^\i}t Fr{\'e}nay. 2021.
\newblock Legal requirements on explainability in machine learning.
\newblock \emph{Artificial Intelligence and Law}, 29(2):149--169.

\bibitem[{Casey and Niblett(2016)}]{casey2016death}
Anthony~J Casey and Anthony Niblett. 2016.
\newblock The death of rules and standards.
\newblock \emph{Ind. LJ}, 92:1401.

\bibitem[{Cui et~al.(2023)Cui, Shen, and Wen}]{cui2023survey}
Junyun Cui, Xiaoyu Shen, and Shaochun Wen. 2023.
\newblock A survey on legal judgment prediction: Datasets, metrics, models and challenges.
\newblock \emph{IEEE Access}, 11:102050--102071.

\bibitem[{Devlin et~al.(2019)Devlin, Chang, Lee, and Toutanova}]{devlin2019bert}
Jacob Devlin, Ming-Wei Chang, Kenton Lee, and Kristina Toutanova. 2019.
\newblock Bert: Pre-training of deep bidirectional transformers for language understanding.
\newblock In \emph{Proceedings of the 2019 conference of the North American chapter of the association for computational linguistics: human language technologies, volume 1 (long and short papers)}, pages 4171--4186.

\bibitem[{Dong and Niu(2021)}]{dong2021rformer}
Qian Dong and Shuzi Niu. 2021.
\newblock Legal judgment prediction via relational learning.
\newblock In \emph{Proceedings of the 44th international ACM SIGIR conference on research and development in information retrieval}, pages 983--992.

\bibitem[{Feng et~al.(2022)Feng, Li, and Ng}]{Feng2022epm}
Yi~Feng, Chuanyi Li, and Vincent Ng. 2022.
\newblock Legal judgment prediction via event extraction with constraints.
\newblock In \emph{Proceedings of the 60th annual meeting of the association for computational linguistics (volume 1: long papers)}, pages 648--664.

\bibitem[{Guha et~al.(2023)Guha, Nyarko, Ho, R{\'e}, Chilton, Chohlas-Wood, Peters, Waldon, Rockmore, Zambrano et~al.}]{guha2023legalbench}
Neel Guha, Julian Nyarko, Daniel Ho, Christopher R{\'e}, Adam Chilton, Alex Chohlas-Wood, Austin Peters, Brandon Waldon, Daniel Rockmore, Diego Zambrano, and 1 others. 2023.
\newblock Legalbench: A collaboratively built benchmark for measuring legal reasoning in large language models.
\newblock \emph{Advances in neural information processing systems}, 36:44123--44279.

\bibitem[{Hong et~al.(2023)Hong, Zhuge, Chen, Zheng, Cheng, Wang, Zhang, Wang, Yau, Lin et~al.}]{hong2023metagpt}
Sirui Hong, Mingchen Zhuge, Jonathan Chen, Xiawu Zheng, Yuheng Cheng, Jinlin Wang, Ceyao Zhang, Zili Wang, Steven Ka~Shing Yau, Zijuan Lin, and 1 others. 2023.
\newblock Metagpt: Meta programming for a multi-agent collaborative framework.
\newblock In \emph{The Twelfth International Conference on Learning Representations}.

\bibitem[{Huang et~al.(2023)Huang, Tao, Zhang, An, Jiang, Chen, Wu, and Feng}]{huang2023lawyer}
Quzhe Huang, Mingxu Tao, Chen Zhang, Zhenwei An, Cong Jiang, Zhibin Chen, Zirui Wu, and Yansong Feng. 2023.
\newblock Lawyer llama technical report.
\newblock \emph{arXiv preprint arXiv:2305.15062}.

\bibitem[{Kaplow(2013)}]{kaplow2013rules}
Louis Kaplow. 2013.
\newblock Rules versus standards: An economic analysis.
\newblock In \emph{Scientific Models of Legal Reasoning}, pages 11--84. Routledge.

\bibitem[{Katz et~al.(2024)Katz, Bommarito, Gao, and Arredondo}]{katz2024gpt}
Daniel~Martin Katz, Michael~James Bommarito, Shang Gao, and Pablo Arredondo. 2024.
\newblock Gpt-4 passes the bar exam.
\newblock \emph{Philosophical Transactions of the Royal Society A}, 382(2270):20230254.

\bibitem[{Kim(2014)}]{kim2014convolutional}
Yoon Kim. 2014.
\newblock \href {https://doi.org/10.3115/v1/D14-1181} {Convolutional neural networks for sentence classification}.
\newblock In \emph{Proceedings of the 2014 Conference on Empirical Methods in Natural Language Processing ({EMNLP})}, pages 1746--1751, Doha, Qatar. Association for Computational Linguistics.

\bibitem[{Liu et~al.(2024)Liu, Feng, Xue, Wang, Wu, Lu, Zhao, Deng, Zhang, Ruan et~al.}]{liu2024deepseek}
Aixin Liu, Bei Feng, Bing Xue, Bingxuan Wang, Bochao Wu, Chengda Lu, Chenggang Zhao, Chengqi Deng, Chenyu Zhang, Chong Ruan, and 1 others. 2024.
\newblock Deepseek-v3 technical report.
\newblock \emph{arXiv preprint arXiv:2412.19437}.

\bibitem[{Liu et~al.(2022)Liu, Du, Li, Pan, and Ming}]{liu2022ctm}
Dugang Liu, Weihao Du, Lei Li, Weike Pan, and Zhong Ming. 2022.
\newblock Augmenting legal judgment prediction with contrastive case relations.
\newblock In \emph{Proceedings of the 29th international conference on computational linguistics}, pages 2658--2667.

\bibitem[{Luo et~al.(2017)Luo, Feng, Xu, Zhang, and Zhao}]{luo2017mlac}
Bingfeng Luo, Yansong Feng, Jianbo Xu, Xiang Zhang, and Dongyan Zhao. 2017.
\newblock \href {https://doi.org/10.18653/v1/D17-1289} {Learning to predict charges for criminal cases with legal basis}.
\newblock In \emph{Proceedings of the 2017 Conference on Empirical Methods in Natural Language Processing}, pages 2727--2736, Copenhagen, Denmark. Association for Computational Linguistics.

\bibitem[{Luo et~al.(2025)Luo, Huang, Jiang, and Feng}]{luo-etal-2025-automating}
Kangcheng Luo, Quzhe Huang, Cong Jiang, and Yansong Feng. 2025.
\newblock \href {https://doi.org/10.18653/v1/2025.acl-long.204} {Automating legal interpretation with {LLM}s: Retrieval, generation, and evaluation}.
\newblock In \emph{Proceedings of the 63rd Annual Meeting of the Association for Computational Linguistics (Volume 1: Long Papers)}, pages 4015--4047, Vienna, Austria. Association for Computational Linguistics.

\bibitem[{Madaan et~al.(2023)Madaan, Tandon, Gupta, Hallinan, Gao, Wiegreffe, Alon, Dziri, Prabhumoye, Yang et~al.}]{madaan2023self}
Aman Madaan, Niket Tandon, Prakhar Gupta, Skyler Hallinan, Luyu Gao, Sarah Wiegreffe, Uri Alon, Nouha Dziri, Shrimai Prabhumoye, Yiming Yang, and 1 others. 2023.
\newblock Self-refine: Iterative refinement with self-feedback.
\newblock \emph{Advances in Neural Information Processing Systems}, 36:46534--46594.

\bibitem[{Qian et~al.(2024)Qian, Liu, Liu, Chen, Dang, Li, Yang, Chen, Su, Cong et~al.}]{qian2024chatdev}
Chen Qian, Wei Liu, Hongzhang Liu, Nuo Chen, Yufan Dang, Jiahao Li, Cheng Yang, Weize Chen, Yusheng Su, Xin Cong, and 1 others. 2024.
\newblock Chatdev: Communicative agents for software development.
\newblock In \emph{Proceedings of the 62nd Annual Meeting of the Association for Computational Linguistics (Volume 1: Long Papers)}, pages 15174--15186.

\bibitem[{Rafailov et~al.(2023)Rafailov, Sharma, Mitchell, Manning, Ermon, and Finn}]{rafailov2023dpo}
Rafael Rafailov, Archit Sharma, Eric Mitchell, Christopher~D Manning, Stefano Ermon, and Chelsea Finn. 2023.
\newblock Direct preference optimization: Your language model is secretly a reward model.
\newblock \emph{Advances in neural information processing systems}, 36:53728--53741.

\bibitem[{Shi et~al.(2025)Shi, Zhu, Ji, Li, Zhang, Zhang, Zhu, Xu, Han, and Guo}]{shi-etal-2025-legalreasoner}
Weijie Shi, Han Zhu, Jiaming Ji, Mengze Li, Jipeng Zhang, Ruiyuan Zhang, Jia Zhu, Jiajie Xu, Sirui Han, and Yike Guo. 2025.
\newblock \href {https://doi.org/10.18653/v1/2025.acl-long.361} {{L}egal{R}easoner: Step-wised verification-correction for legal judgment reasoning}.
\newblock In \emph{Proceedings of the 63rd Annual Meeting of the Association for Computational Linguistics (Volume 1: Long Papers)}, pages 7297--7313, Vienna, Austria. Association for Computational Linguistics.

\bibitem[{Shinn et~al.(2023)Shinn, Cassano, Gopinath, Narasimhan, and Yao}]{shinn2023reflexion}
Noah Shinn, Federico Cassano, Ashwin Gopinath, Karthik Narasimhan, and Shunyu Yao. 2023.
\newblock Reflexion: Language agents with verbal reinforcement learning.
\newblock \emph{Advances in Neural Information Processing Systems}, 36:8634--8652.

\bibitem[{{The Supreme People's Court of the PRC}(2024)}]{spc2024report}
{The Supreme People's Court of the PRC}. 2024.
\newblock Work report of the supreme people's court.
\newblock Retrieved from \url{http://www.court.gov.cn}.

\bibitem[{Wei et~al.(2022)Wei, Wang, Schuurmans, Bosma, Xia, Chi, Le, Zhou et~al.}]{wei2022chain}
Jason Wei, Xuezhi Wang, Dale Schuurmans, Maarten Bosma, Fei Xia, Ed~Chi, Quoc~V Le, Denny Zhou, and 1 others. 2022.
\newblock Chain-of-thought prompting elicits reasoning in large language models.
\newblock \emph{Advances in neural information processing systems}, 35:24824--24837.

\bibitem[{Wu et~al.(2024)Wu, Bansal, Zhang, Wu, Li, Zhu, Jiang, Zhang, Zhang, Liu et~al.}]{wu2024autogen}
Qingyun Wu, Gagan Bansal, Jieyu Zhang, Yiran Wu, Beibin Li, Erkang Zhu, Li~Jiang, Xiaoyun Zhang, Shaokun Zhang, Jiale Liu, and 1 others. 2024.
\newblock Autogen: Enabling next-gen llm applications via multi-agent conversations.
\newblock In \emph{First Conference on Language Modeling}.

\bibitem[{Wu et~al.(2023)Wu, Zhou, Liu, Lu, Liu, Zhang, Sun, Wu, and Kuang}]{wu2023pljp}
Yiquan Wu, Siying Zhou, Yifei Liu, Weiming Lu, Xiaozhong Liu, Yating Zhang, Changlong Sun, Fei Wu, and Kun Kuang. 2023.
\newblock \href {https://doi.org/10.18653/v1/2023.emnlp-main.740} {Precedent-enhanced legal judgment prediction with {LLM} and domain-model collaboration}.
\newblock In \emph{Proceedings of the 2023 Conference on Empirical Methods in Natural Language Processing}, pages 12060--12075, Singapore. Association for Computational Linguistics.

\bibitem[{Xiao et~al.(2018)Xiao, Zhong, Guo, Tu, Liu, Sun, Feng, Han, Hu, Wang et~al.}]{xiaocail2018}
Chaojun Xiao, Haoxi Zhong, Zhipeng Guo, Cunchao Tu, Zhiyuan Liu, Maosong Sun, Yansong Feng, Xianpei Han, Zhen Hu, Heng Wang, and 1 others. 2018.
\newblock Cail2018: A large-scale legal dataset for judgment prediction.
\newblock \emph{arXiv preprint arXiv:1807.02478}.

\bibitem[{Xu et~al.(2020)Xu, Wang, Chen, Pan, Wang, and Zhao}]{xu2020ladan}
Nuo Xu, Pinghui Wang, Long Chen, Li~Pan, Xiaoyan Wang, and Junzhou Zhao. 2020.
\newblock \href {https://doi.org/10.18653/v1/2020.acl-main.280} {Distinguish confusing law articles for legal judgment prediction}.
\newblock In \emph{Proceedings of the 58th Annual Meeting of the Association for Computational Linguistics}, pages 3086--3095, Online. Association for Computational Linguistics.

\bibitem[{Yang et~al.(2019)Yang, Jia, Zhou, and Luo}]{yang2019mpbfn}
Wenmian Yang, Weijia Jia, Xiaojie Zhou, and Yutao Luo. 2019.
\newblock Legal judgment prediction via multi-perspective bi-feedback network.
\newblock In \emph{Proceedings of the 28th International Joint Conference on Artificial Intelligence}, IJCAI'19, page 4085–4091. AAAI Press.

\bibitem[{Yang et~al.(2025)Yang, Chai, Shao, Song, Qi, Rui, and Zhang}]{yang2025agentnet}
Yingxuan Yang, Huacan Chai, Shuai Shao, Yuanyi Song, Siyuan Qi, Renting Rui, and Weinan Zhang. 2025.
\newblock Agentnet: Decentralized evolutionary coordination for llm-based multi-agent systems.
\newblock \emph{arXiv preprint arXiv:2504.00587}.

\bibitem[{Yue et~al.(2021)Yue, Liu, Jin, Wu, Zhang, An, Cheng, Yin, and Wu}]{yue2021neurjudge}
Linan Yue, Qi~Liu, Binbin Jin, Han Wu, Kai Zhang, Yanqing An, Mingyue Cheng, Biao Yin, and Dayong Wu. 2021.
\newblock Neurjudge: A circumstance-aware neural framework for legal judgment prediction.
\newblock In \emph{Proceedings of the 44th international ACM SIGIR conference on research and development in information retrieval}, pages 973--982.

\bibitem[{Zhang et~al.(2025)Zhang, Fu, Wan, Yu, Wang, and Yan}]{zhang2025gmemory}
Guibin Zhang, Muxin Fu, Guancheng Wan, Miao Yu, Kun Wang, and Shuicheng Yan. 2025.
\newblock G-memory: Tracing hierarchical memory for multi-agent systems.
\newblock \emph{arXiv preprint arXiv:2506.07398}.

\bibitem[{Zhong et~al.(2018)Zhong, Guo, Tu, Xiao, Liu, and Sun}]{zhong2018topjudge}
Haoxi Zhong, Zhipeng Guo, Cunchao Tu, Chaojun Xiao, Zhiyuan Liu, and Maosong Sun. 2018.
\newblock Legal judgment prediction via topological learning.
\newblock In \emph{Proceedings of the 2018 conference on empirical methods in natural language processing}, pages 3540--3549.

\end{thebibliography}
